\documentclass[10pt,final,twocolumn,twoside]{IEEEtran}
\usepackage{framed}
\usepackage{graphicx} \usepackage{amsmath} \usepackage{amssymb}
\usepackage{color}
\usepackage{amssymb}
\usepackage{amsthm} 
\usepackage[linesnumbered,ruled,vlined]{algorithm2e}
\SetKwComment{Comment}{$\triangleright$\ }{}
\newlength\algowd

\usepackage{cite} 
\usepackage[latin1]{inputenc}
\usepackage[caption=false,font=footnotesize]{subfig} 
\usepackage{stfloats}
\usepackage{url}
\usepackage{mathtools}
\usepackage{varwidth}

\usepackage{xpatch}
\makeatletter
\patchcmd\@makecaption{\\}{.~}{}{\fail}
\makeatletter
\newcommand{\mb}[1]{\mbox{#1}}

\newcommand{\PPUSCH}{P_{\mbox{\scriptsize PUSCH}}}
\newcommand{\PCMAX}{P_{\mbox{\scriptsize CMAX}}}
\newcommand{\MPUSCH}{M_{\mbox{\scriptsize PUSCH}}}
\newcommand{\DTF}{\Delta_{\mbox{\scriptsize TF}}}
\newcommand{\PNOM}{P_{\mbox{\scriptsize NOMINAL}}}
\newcommand{\PUE}{P_{\mbox{\scriptsize UE}}}
\newcommand{\PLN}{\Delta PL_{\mbox{\scriptsize n}}}
\newcommand{\PL}[1]{\Delta PL_{\mbox{\scriptsize #1}}}
\newcommand{\DPUSCH}{\delta_{\mbox{\scriptsize PUSCH}}}
\newcommand{\KPUSCH}{K_{\mbox{\scriptsize PUSCH}}}
\newcommand{\PAER}{P_{\mbox{\scriptsize 0A}}}
\newcommand{\PTERR}{P_{\mbox{\scriptsize 0T}}}
\begin{document}

\title{Interference Mitigation Methods for Unmanned Aerial Vehicles Served by Cellular Networks}
\date{\today}
\author{Vijaya~Yajnanarayana, Y.-P. Eric Wang, Shiwei~Gao, Siva~Muruganathan, Xingqin~Lin\\
  Ericsson\\
  Email: vijaya.yajnanarayana@ericsson.com}
\maketitle

\begin{abstract}
A main challenge in providing connectivity to the low altitude unmanned aerial vehicles (UAVs) through existing cellular network arises due to the increased interference in the network. The increased altitude and favourable propagation condition cause UAVs to generate more interference to the neighbouring cells, and at the same time experience more interference from the downlink transmissions of the neighbouring base stations. The uplink interference problem may result in terrestrial UEs having degraded performance, whereas the downlink interference problem may make it challenging for a UAV to maintain connection with the network. In this paper, we propose several uplink and downlink interference mitigation techniques to address these issues. The results indicate that the proposed solutions can reduce the uplink throughput degradation of terrestrial UEs and ensure UAVs to remain in LTE coverage under the worst case scenarios when all the base stations transmit at full power.  \\

\end{abstract}
%\vspace{-0.5in}
\section{Introduction}
\label{sec:intro}

Offering broadband connectivity to the low altitude, small unmanned aerial vehicles (UAVs) is an important emerging field. UAVs find applications in a wide variety of industries and services including package delivery, agriculture, surveillance, search and rescue. To safely control, track, and manage the growing fleet of UAVs, it is essential to develop innovative communication technologies that support wide-area, beyond visual line-of-sight (LOS) communication \cite{FAA-spec}. The 5th generation mobile networks (5G), including both New Radio (NR) and further Long-Term Evolution (LTE) advancement, will address a variety of applications from eMBB (enhanced Mobile Broadband) to mMTC (massive Machine Type Communications) to URLLC (Ultra-Reliable and Low Latency Communications). With advanced capabilities, 5G has the potential to provide reliable, secure, high throughput, low latency mobile broadband connectivity to the UAVs. However, there exist challenges in providing ubiquitous coverage for UAVs using cellular networks. The 3rd Generation Partnership Project (3GPP) has recently concluded a study item on enhanced LTE support for aerial vehicles\cite{TR36777}. One of the key findings of this study is that the introduction of aerial vehicles into the existing LTE networks will increase interference. In the study item, interference mitigation solutions requiring standard enhancements and those that are purely implementation based were also identified.

\begin{figure}[t]
  \centering
  \includegraphics[width=3.2 in]{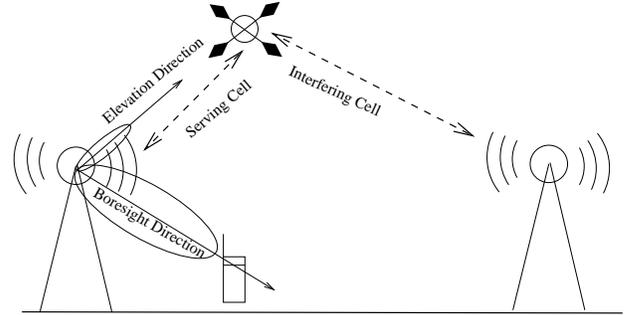}
  \caption{Illustration of connectivity for small UAVs with wireless cellular networks.}
  \label{fig:tilt}
\end{figure}

In a cellular network, in order to focus the energy to the terrestrial users as well as to curtail the interference to the neighbouring cells, the base station (BS) antennas are typically down-tilted. This has a profound implication on the path gain and signal-to-interference-plus-noise (SINR) for the UAVs. Due to the down-tilt of the BS antennas, the UAVs would be served by the side-lobes (refer to Fig.~\ref{fig:tilt}) with a reduced antenna gain. However, it is shown in \cite{UAV-mag1} that the benign propagation environment between BS and UAV can compensate for this reduction. Even though the path gain is compensated by the good propagation condition, the interference is still an issue. The increased altitude and favourable propagation condition in the sky result in the UAVs generating more uplink (UL) interference to the neighbouring cells. The increased uplink interference may degrade terrestrial UE performance. In the downlink, the UAVs are victims of interference as signals from many neighbouring BSs may reach the UAVs with strong power levels. A UAV thus may experience degraded SINR, which could be at a level that essential LTE signals can not be received reliably for the UAV to maintain connectivity with the network. In this paper, we present our studies of these problems. In Section~\ref{sec:im}, we illustrate the UAV interference problems with system-level simulation results and give an overview of the existing interference mitigation methods. Then in Section~\ref{sec:mpc} and Section~\ref{sec:coverage}, we discuss uplink and downlink interefernce mitigation solutions. Simulation results are presented in Section~\ref{sec:sim}, and finally we conclude with key observations in Section~\ref{sec:conclusion}.

%\pagebreak

\section{UAV Interference Problems and Interference Mitigation}
\label{sec:im}

\begin{figure}[t]
  \centering
  \includegraphics[width=3.2 in]{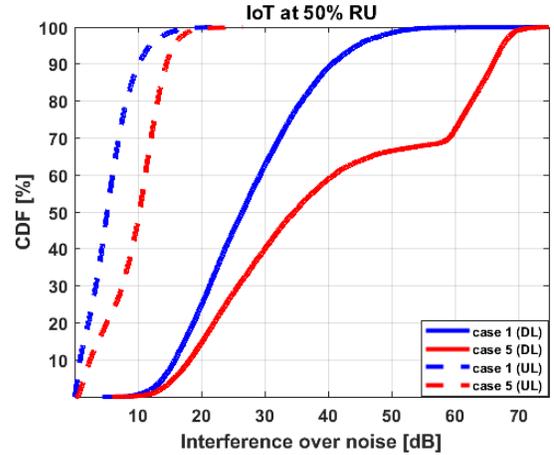}
  \caption{Interference over thermal noise (IoT) statistics for UL and DL at 50\% resource utilization (RU) using the configuration setup defined in \cite{TR36777}. As defined in \cite{TR36777}, case-1 has only terrestrial UEs and case-5 is a hybrid deployment with 5 UAVs per cell.}
  \label{fig:iot}
\end{figure}

To illustrate UAV interference problem in the UL an DL, we show how interference level is increased when UAVs connect to the network compared to when only terrestrial UEs are in the network. Example results are shown in Fig.~\ref{fig:iot}. The blue curves represent interference over thermal noise (IoT) statistics when the network serves only terrestrial UEs. The red curves represent IoT statistics when the network serves both terrestrial and UAVs according to a scenario defined in \cite{TR36777}, where there are 10 terrestrial UEs and 5 UAVs per cell. The statistics are collected from all UEs in the network with a traffic load set to result in approximately 50\% resource utilization (RU) in the network. It can be seen that both UL and DL interference levels are raised when UAVs are served.

There exists a rich set of tools in terms of both standards and implementation that have been studied and developed to deal with interference. One solution may be deploying dedicated cells for serving the UAVs, where the antenna patterns of the BSs are pointed towards to the sky instead of down-tilted. These dedicated cells will be particularly helpful in drone hotspots where frequent and dense drone takeoffs and landings occur. Deploying such {\it air cells} however requires further investment and this approach will become a more economically attractive solution when the number of UAVs connecting to the network increases. Another possible solution is beamforming. Considering the size of the UAVs it is possible to equip the UAVs with multiple antennas which can be used to beam-form the uplink signal to serving cell to reduce interference to the neighbouring cells. Similarly, in the downlink, a UAV can steer its receiving beam pattern to its serving BS to reduce interference from neighbouring cells. While this solution is effective toward the serving cell, it may not be sufficient in handover scenarios when the UE needs synchronize to and perform measurements on neighbouring cells to identify potential handover targets.

%One prominent interference mitigation tool is coordinated multipoint (CoMP) transmission and reception (and its variants). The new challenge here is that UAVs receive interfering signals from %more ground BSs in the downlink and their uplink signals are visible to more cells due to more LOS propagation conditions. Thus, the methods need to scale for a large set of cells without much %complications due to the requirements on additional pilots, synchronization, scheduling, etc.
Minimizing the impact on the performance of terrestrial UEs is critical for the mobile network operators to be able to reuse existing deployed terrestrial networks for also serving UAVs. In the below sections, we present solutions that allow the mobile network operators to achieve these goals. For the UL interference problem, we propose
adaptations to the existing power control methods to alleviate UAV interference to terrestrial UEs. For the downlink interference problem, 
we consider the 3GPP LTE Release~13 coverage extension features developed for machine type communications (MTC) to ensure an aerial UE is in coverage even under the worst case interference scenario. 

\section{UL Interference Mitigation}
\label{sec:mpc}
The power control mechanism ensures that the transmit power of different uplink channels are controlled such that these channels are received at the BSs at appropriate power level. The power control procedure aims to control the received power to be just enough to demodulate the channel (target received power), at the same time the transmit power at UEs are not unnecessarily high as it could create interference to the other uplink transmissions \cite{LTE-book}. In many standards like LTE, the transmit power of the UE depends on the DL pathloss and target received power at the serving BS. Using the LTE Physical Uplink Shared Channel (PUSCH) as an example, a typical UE's UL power at subframe $i$ can be expressed as

\begin{equation}
  \label{eq:pc}
  \PPUSCH(i) = \min \left\{\begin{array}{l}
                     \PCMAX,\\
                             \left(
                             \begin{array}{l}
                             10\log_{10}(\MPUSCH(i)) \\
                               \quad +P_0+\alpha PL \\ \quad \quad  + \DTF(i)  + f(i)
                               \end{array} \right)
                   \end{array}\right\},
\end{equation}
Where
\begin{itemize}
\item $\PCMAX$ is the configured maximum UE transmit power in dBm
\item $\MPUSCH(i)$ is the bandwidth of the PUSCH resource assignment expressed in number of resource blocks valid for subframe $i$
\item $P_0$ is an open loop power control parameter in dBm composed of the sum of a cell specific component, $\PNOM$ and a UE specific component,   $\PUE$

\item $\alpha \in  \left\{0,0.4,0.5,0.6,0.7,0.8,0.9,1\right\} $ is a fractional path loss compensation power control parameter
\item  $PL$ is the downlink pathloss estimate computed at the UE in dB
\item $\DTF (i)$  is  an offset which can be used to ensure that the received SINR matches the SINR required for a given modulation and coding scheme (MCS) selected by the eNB
\item $f(i)$ is the closed loop power control adjustment   
\end{itemize}

\subsection{Open Loop Power Control}
\label{sec:opc}
Power control equation shown in \eqref{eq:pc} has two parts, an open loop part consisting of $10\log_{10}(\MPUSCH(i)) +P_0+\alpha PL + \DTF(i)$  and closed loop part consisting of $f(i)$. As discussed in Section~\ref{sec:intro}, due to the line-of-sight propagation  condition to the neighbouring cells, UAVs can cause significant UL interference if not properly managed (refer to Fig~\ref{fig:iot}). One way to reduce the UL interference caused by UAVs is to use smaller value of $P_0$ and/or $\alpha$. As mentioned above $P_0$ consists of a cell specific part, $\PNOM$ and UE specific part, $\PUE$. The UE specific parameter can be tuned to adjust (reduce) the transmit power of the UAVs.

Another approach could be to add an additional power adjustment factor at BS, $\beta$, to the computed power based on the pathlosses to neighbour cells. For example, the transmit power can be adjusted by the BS based on the ratio between the serving cell pathloss and the $n$-th strongest neighbour cell pathloss, denoted as $\PLN$. If $\PLN$ is low, then $\beta$ can be adjusted to reduce the UL transmit power to reduce interference to neighbouring cells. An example is shown in Table \ref{tab:beta}, where pathloss ratio to the third strongest neighbour cell is considered.

\begin{table}[t]
   \caption{An example of power adjustment based on $\PLN$}
	\centering
	\vspace{-0.16in}
	\begin{tabular}[t]{|c|c|}
          \hline
          $\PL{3}$ (dB) & Additional Power Adjustment ($\beta$) (dB) \\
          \hline
          $0$ &	$-6$ \\
          $1$ &	$-5$ \\
          $2$ &	$-4$ \\
          $3$ &	$-3$ \\
          $4$ &	$-2$ \\
          $5$ &	$-1$ \\
          $>5$ & $0$ \\\hline
	\end{tabular}
        \label{tab:beta}
\end{table}

For $\PLN$ calculation, the transmit power of the neighbour cell is required at a UE. Alternatively, the ratio between the serving cell RSRP and the $n$-th neighbour cell RSRP may be used instead of $\PLN$.

\subsection{Closed Loop Power Control}
\label{sec:cpc}
The closed loop power control part is represented by $f(i)$ in \eqref{eq:pc}. The $f(i)=f(i-1)+\DPUSCH(i-\KPUSCH)$, if accumulation is allowed, else  $f(i)=\DPUSCH(i-\KPUSCH)$. The variable $\DPUSCH(i-\KPUSCH)$ denotes the transmit power control (TPC) command signaled to the UE at subframe $(i-\KPUSCH)$.  For  convenience, TPC command, $\DPUSCH(i-\KPUSCH)$, will be henceforth referred to as $\DPUSCH$.
  
% \begin{figure}[b]
%   \centering
%   \includegraphics[width=2.5 in]{CPC}
%   \caption{Illustration of the close loop power control.}
%   \label{fig:cpc}
% \end{figure}

The transmit power at UE, $\PPUSCH$, may be adjusted such that the received power at BS is equal to a predetermined target receive power. One or more TPC commands are sent to the UE by the BS to adjust its transmit power, $\PPUSCH$, to achieve the target received power at BS. Since UAVs are the main cause of interference, significant gains can be achieved by  setting the target receive power individually based on the serving cell RSRP and the neighbour cell RSRP for UAVs, while keeping it same for all terrestrial UEs. 

\begin{figure}[b]
  \centering
  \includegraphics[width=3.5 in]{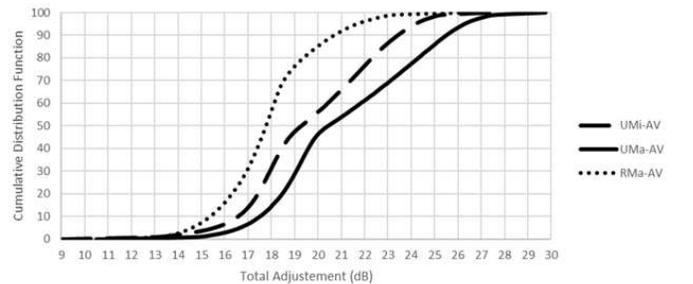}
  \caption{Cumulative distribution function of the total adjustment ($P_{cpc}$) needed in closed loop power control for three UAV scenarios.}
  \label{fig:comp}
\end{figure}

In typical cellular networks, the open-loop power control compensates for the fractional path-loss (depending on choice of $\alpha$) and the remaining path-loss is compensated based on the closed loop power control. Hence the total adjustment using closed loop power control is given by $P_{cpc}=(1-\alpha)PL$. Three propagation models for UAVs are specified in \cite{TR36777}. Based on these propagation models, path losses for the UMi-AV (urban micro with aerial vehicles), UMa-AV (urban macro with aerial vehicles) and RMa-AV (rural macro with aerial vehicles) scenarios defined in \cite{TR36777} can be generated. Figure ~\ref{fig:comp} shows the cumulative distribution function of the total adjustments needed at the UAVs for these three scenarios when the fractional path loss compensation factor $\alpha$ is set to $0.8$.

For the existing closed loop power control scheme in \cite{TS36213}, the range of values currently supported for $P_0$ are too limited for the UAV scenarios.  It should be noted that the closed loop power control for UAVs also needs to cope with potential fast signal change in the sky (since UAVs may be served by the side-lobes of BS antennas (refer to Fig.~\ref{fig:tilt})). Hence, specification enhancements for increased step size of $\DPUSCH$  is needed to cover large power adjustments (refer to Fig.~\ref{fig:comp}) in short amount of time.

\section{DL Interference Mitigation}
\label{sec:coverage}

The power control based technique presented in the previous section mainly addresses the uplink interference problems. In the downlink, due to favorable propagation condition, an UAV may experience strong inter-cell interference. A number of downlink physical channels, e.g. Synchronization Channel (SCH) and Physical Broadcast Channel (PBCH), are used for cell acquisition \cite{LTE-book}. These channels are broadcast in nature as they need to be received by all UEs in the cell, and thus UE-specific beamforming or power control cannot be applied. For UAVs, the SINR for these channels may become lower than the normal coverage that an LTE network aims to achieve. Coverage extension (CE) is a feature set introduced in LTE Release 13 \cite{CIoT-book}. It was motivated for supporting machine type communications (MTC), which have stringent requirements on coverage. The LTE CE features however are not limited to MTC devices. For example, it can help UAVs complete cell acquisitions successfully in the presence of significant inter-cell interference. During cell acquisition, the UE needs to detect SCH, get master information through PBCH and other system information through Physical Downlink Shared Channel (PDSCH). Cell acquisition is an essential step for supporting handover procedures. LTE coverage extension targets 155.7 dB maximum coupling loss (MCL), and the link budget for all the DL physical channels needed to achieve this MCL during cell acquisition is shown in Table \ref{tab:CElinkbudget}. Rel-13 coverage extension is achieved mainly through repetitions. Repetitions give rise to higher signal energy which extends the coverage. Repetitions also help mitigate interference through a processing gain over the interfering signals. As CE achieves 155.7 dB MCL, it is guaranteed that UEs will have SINR higher than $-14.3$ dB and will be able to achieve cell acquisition, as shown in Table~\ref{tab:CElinkbudget}.

 \begin{table}[t]
	\caption{LTE coverage extension link budget}
	\centering
	\vspace{-0.16in}
	\begin{tabular}[t]{|c|c|c|c|}
		\hline
		Physical channel & SCH & PBCH & PDSCH \\
		\hline\hline
		Max Tx power per LTE carrier  (dBm)	& 46.0 &	46.0 &	46.0\\\hline
		Actual Tx power (dBm) &	32.0 &	36.8 &	36.8\\
		(adjusted for signal bandwidth) & & & \\\hline
		Thermal noise density (dBm/Hz)	&-174.0 &	-174.0 &	-174.0\\\hline
		Receiver noise figure (dB)	&9	&9	&9\\\hline
		Occupied channel bandwidth (Hz)	&360000	&1080000	&1080000\\\hline
		Effective noise power (dBm) & -109.4 & -104.7 & -104.7\\\hline
		Required SINR (dB)	&{\bf-14.3} & {\bf -14.2} &	{\bf -14.2} \\\hline
		Receiver sensitivity (dBm)	& 	-123.7 	& -118.9  & -118.9 
		 \\\hline
		MCL (dB) & 155.7 & 155.7 & 155.7 \\\hline	
	\end{tabular}
	\label{tab:CElinkbudget}
\end{table}

\section{Simulation Results}
\label{sec:sim}

We first present the results for the open loop power control method discussed in Section \ref{sec:opc}.  Here we configure the UAVs  and terrestrial UEs with different $P_0$ values denoted as $\PAER$ and $\PTERR$ respectively. We evaluate the performance for six combinations shown in the Table \ref{tab:p0-sweep}. For simulations, we have used the urban macro (UMa) scenario with $\mb{case-5}$ deployment (5 UAVs and 10 terrestrial UEs per cell) as discussed in \cite{TR36777}. Combination-1 (refer to Table~\ref{tab:p0-sweep})  is used as a baseline, since, here the power control parameters for UAVs are same as in terrestrial UEs. \footnote{ In Combination-1, from power control perspective,  UAVs are treated as though they are terrestrial UEs.}

\begin{figure}[t]
  \centering
  \includegraphics[width=3.2 in]{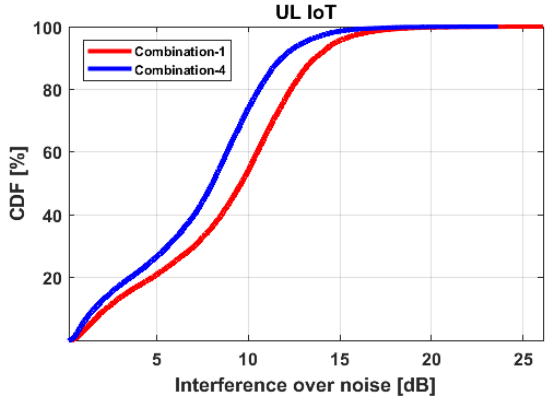}
  \caption{Cumulative distribution function of the UL IoT considering all UEs (both UAV and terrestrial) in the system at 50\% RU. (5 UAVs and 10 terrestrial UEs per cell)}
  \label{fig:uliot}
\end{figure}

\begin{figure}[h]
  \centering
  \includegraphics[width=3 in]{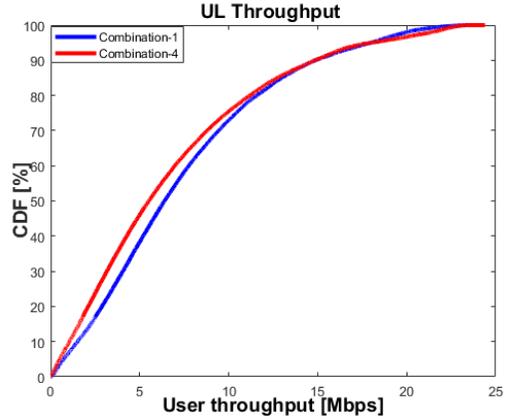}
  \caption{The UL throughput for case-5 UMa-AV scenario considering all UEs in the system at 50\% RU.}
  \label{fig:ulthroughputcdf}
\end{figure}

Lowering the  $\PAER$ value reduces the transmit power of the UAVs thus reducing the overall interference of the system. Figure~\ref{fig:uliot} shows the decrease in the interference level when $\PAER$ is reduced from -$85$ to -$88$~dBm. The throughput of the system (considering both UAVs and terrestrial UEs) is shown in Fig.~\ref{fig:ulthroughputcdf}. Notice an approximate gain of  $0.8$ and $1$~Mbps at $20$ and $50$ percentile numbers respectively by reducing $\PAER$ from  -$85$ to -$88$~dBm. 

\begin{figure}[t]
  \fbox{
  \centering
  \includegraphics[width=3 in]{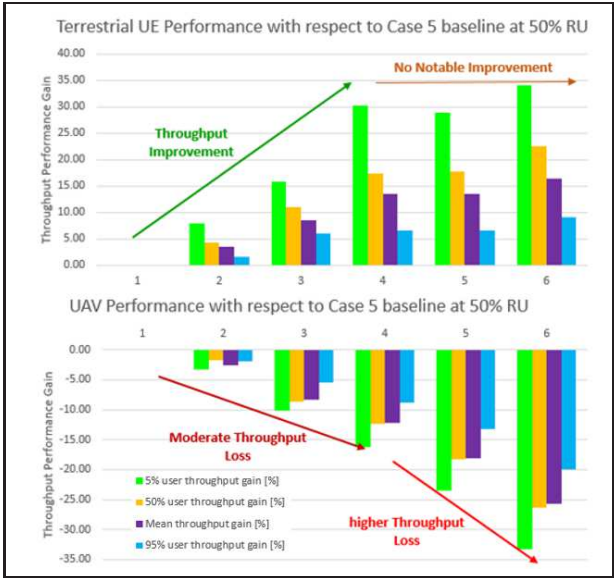}
  }
  \caption{Uplink terrestrial UE throughput results with different $P_0$ values for UAVs and terrestrial UEs with Combination-1 in Table \ref{tab:p0-sweep} as the baseline with 50\% resource utilization (RU).}
  \label{fig:opcr}
\end{figure}

The terrestrial and UAV uplink throughput gains are shown in Fig.~\ref{fig:opcr}. Note that the combination number is shown in the horizontal axis. From Fig.~\ref{fig:opcr}, it is evident that the terrestrial UE throughput performance can be improved by configuring the UAVs with a lower $\PAER$  value. The achieved improvement is significant as it can help in protecting the terrestrial users' throughput from increased interference from UAVs. Also, notice that the decrease in transmit power of UAVs reduces the throughput of the UAVs. When $\PAER$  is gradually reduced from $-85$~dBm to $-88$~dBm, moderate UAV throughput losses is observed.  However, reducing $\PAER$ even further results in higher UAV throughput losses without significant improvement to the terrestrial users.

\begin{table}[t]
  \caption{The $P_0$ values for Terrestrial UEs and UAVs}
  \centering
  \vspace{-0.16in}
  \begin{tabular}[t]{|c||c|c|}
    \hline
    Combination & $\PTERR$ [dBm] & $\PAER$  [dBm]\\
    \hline
    1 & -85 & -85 \\
    2 & -85 & -86 \\
    3 & -85 & -87 \\
    4 & -85 & -88 \\
    5 & -85 & -89 \\
    6 & -85 & -90 \\
    \hline    
  \end{tabular}
  \label{tab:p0-sweep}
\end{table}

\begin{figure}[t]
  \fbox{
  \centering
  \includegraphics[width=3 in]{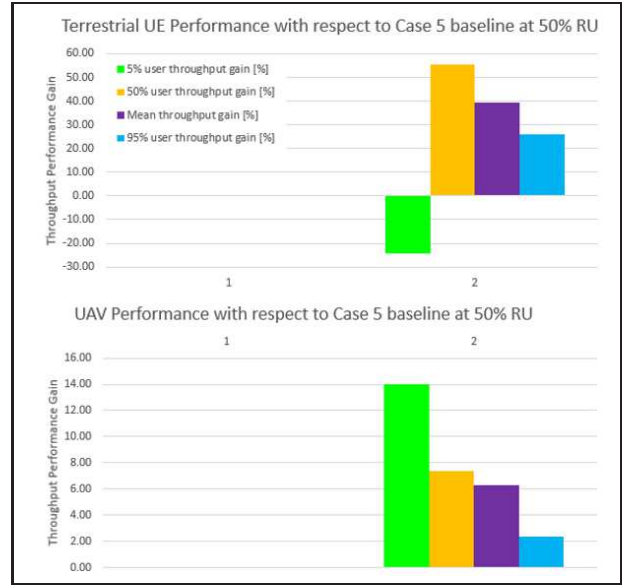}
  }
  \caption{Uplink throughput results for terrestrial and UAVs at 50\% resource utilization (RU).}
   \label{fig:cpcr}
\end{figure}

For the closed loop simulations, we use the same deployment as in the open loop case (urban macro, case-5) and  evaluate the performance for the two combinations described in Table~\ref{tab:cpcr}. In the first combination, the closed loop parameters are kept the same for both terrestrial and UAVs. We use this as the baseline. Combination-2 uses the method described in the Section \ref{sec:cpc}, where, for the UAVs, target receive power is individually adjusted based on the serving cell RSRP and the third strongest neighbour cell RSRP. Results are summarized in Fig.~\ref{fig:cpcr}. From the result, it is evident that terrestrial UEs' and UAVs' mean, $50$ percentile, and $95$ percentile throughput performance are improved by the proposed solution.  However, some cell edge throughput losses are seen for terrestrial UEs.  One reason for this could be due to power limited terrestrial UEs not being able to reach the target received power ($P_0 =-94$~dBm).

\begin{table}[t]
  \caption{The closed loop setup}
  \centering
  \vspace{-0.16in}
  \begin{tabular}[t]{|p{{1 in}}||p{2 in}|}
    \hline
    Combination & Description\\
    \hline
    1 & $\PTERR=\PAER=-85$~dBm \\ \hline 
    2 & $\PTERR = -85$~dBm and $\PAER$ adjusted individually based on serving cell RSRP and neighbour cell RSRP\\
    \hline    
  \end{tabular}
  \label{tab:cpcr}
  
\end{table}
        
\begin{figure}[t]
		\centering
		\includegraphics[width=3 in]{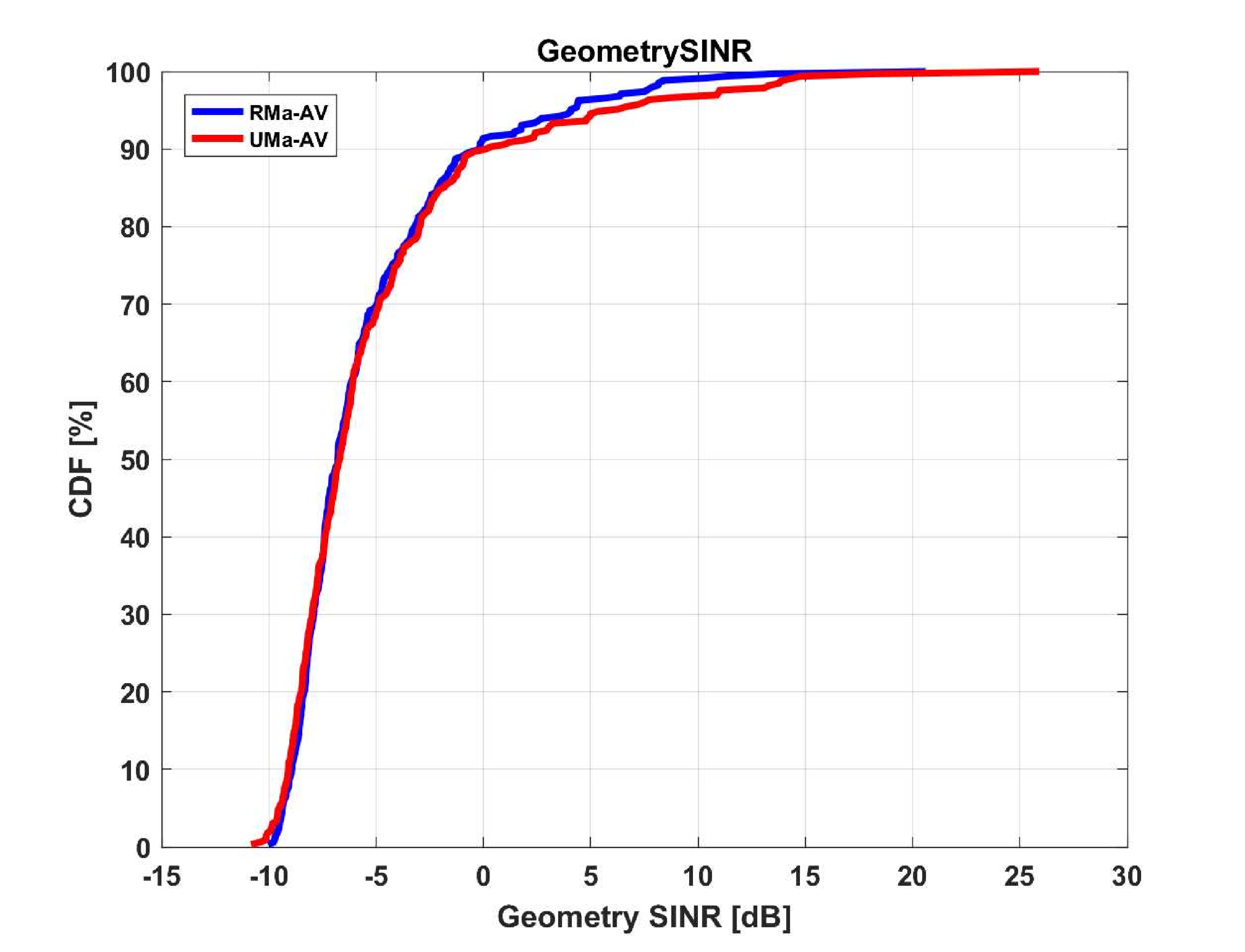}
	\caption{CDF of downlink geometry SINR experienced by UAVs.}
	\label{fig:DL_GSINR}
\end{figure}

Next, we study how DL interference may affect the performance of synchronization and cell acquisition experienced by UAVs. For physical channels carrying unicast traffic, the level of inter-cell interference depends on traffic load. However, for the synchronization and cell acquisition channels, the interference may not depend on the traffic load. One example for such is a system frame number (SFN) synchronized network, which is a possible network configuration when LTE features such as multicast-broadcast single-frequency network (MBSFN) \cite{LTE-book} and Positioning Reference Signal (PRS) \cite{LTE-book} are supported. In such scenarios, all the BSs may transmit SCH and PBCH at the same time. We study the downlink wideband SINR statistics experienced by UAVs. To assess UAVs performance in the worst-case scenarios, the downlink wideband SINR statistics are obtained by assuming all the BSs in the network transmit at full power. We will also refer to this as DL geometry SINR. Figures \ref{fig:DL_GSINR} shows DL geometry SINR collected from UAVs for both RMa-AV and UMa-AV scenarios. Here, we see that the DL geometry SINR can be as low as $-10.8$ dB. However, this SINR value is still higher than the required SINR according to the LTE CE link budget shown in Table \ref{tab:CElinkbudget}. 

\begin{table}[b]
	\caption{probability of synchronization and cell acquisition channels for UAVs}
	\centering
	\vspace{-0.16in}
	\begin{tabular}[t]{|c|c|c|c|c|}
		\hline
		& \multicolumn{2}{c|}{RMa-AV } & \multicolumn{2}{c|}{UMa-AV } \\\cline{2-5}
		& w/o CE & w/CE & w/o CE & w/CE \\\hline
		SCH outage	& 33\% & 0\% & 33\% & 0\% \\\hline
		PBCH outage	& 37\% & 0\% & 38\%	& 0\% \\\hline
		System information outage &	77\% & 0\% & 75\% & 0\% \\\hline
	\end{tabular}
	\label{tab:outage}
\end{table}

In Table \ref{tab:outage}, we summarize the outage probability of synchronization and cell acquisition channels among UAVs, where outage is defined as the received DL SINR for the corresponding channel falling below the respective required SINR threshold. The required SINR threshold for LTE normal coverage is based on the link budget in \cite{TR36888}.

\section{Conclusion and Discussion}
\label{sec:conclusion}

One of the main challenges of reusing terrestrial cellular network for UAV connectivity is interference. In the uplink, UAVs are aggressors, which may generate  interference to neighboring cells due to potentially line-of-sight propagation conditions to many base stations. We showed that existing power control framework can be extended to mitigate uplink interference. Results show that for a scenario with mixed terrestrial UEs and UAVs as defined in 3GPP, the throughput gains for the terrestrial UEs can be  improved by $30\%$ to $50\%$ using the proposed modifications. In the downlink, UAVs are victims of inter-cell interference. To maintain connectivity with the network, an UAV needs to be able to complete cell acquisition reliably so that potential handover target cells can be identified. LTE coverage extension solutions can be adopted to improve cell acquisition for UAVs. By using these tools, the outage probability of cell acquisition channels, e.g. SCH, PBCH, and PDSCH carrying system information, can be reduced to $0\%$ from $33\%$-$75\%$.

\bibliography{main}
\bibliographystyle{IEEE}
 
\end{document}